\documentclass[amssymb,amsfonts,amsmath,reprint,prapplied,floatfix]{revtex4-1}
\usepackage{graphicx}
\usepackage{hyperref}

\begin{document}

\title{Frequency Fluctuations in Tunable and Nonlinear Microwave Cavities}
\date{\today}

\author{B. L. Brock}
\affiliation{Department of Physics and Astronomy, Dartmouth College, Hanover, New Hampshire 03755, USA}

\author{M. P. Blencowe}
\affiliation{Department of Physics and Astronomy, Dartmouth College, Hanover, New Hampshire 03755, USA} 

\author{A. J. Rimberg}
\email{ajrimberg@dartmouth.edu}
\affiliation{Department of Physics and Astronomy, Dartmouth College, Hanover, New Hampshire 03755, USA}

\begin{abstract}
We present a model for how frequency fluctuations comparable to the total cavity linewidth may arise in tunable and nonlinear microwave cavities, and how these fluctuations affect the measurement of scattering matrix elements.  Applying this model to the specific case of a two-sided cavity, we obtain closed-form expressions for the average scattering matrix elements in several important cases.  A key signature of our model is the subtle deformation of the trajectories swept out by scattering matrix elements in the complex plane.  Despite this signature, the fluctuating and non-fluctuating models are qualitatively similar enough to be mistaken for one another.  In the case of tunable cavities we show that if one fails to account for these fluctuations then one will find damping rates that appear to depend on the tuning parameter, which is a common observation in such systems.  In the case of a Kerr cavity, we show that there exists a fundamental lower bound to the scale of these frequency fluctuations in the steady state, imposed by quantum mechanical uncertainty, which can appreciably affect the apparent damping rates of the cavity as the strength of the nonlinearity approaches the cavity linewidth.  By using the model we present as a fitting function for experimental data, however, one can extract both the true damping rates of the cavity and the effective scale of these frequency fluctuations over the scattering measurement's bandwidth.  Lastly, we compare this new method for observing frequency fluctuations to other methods, one of which we extend beyond the regime of small fluctuations, and we discuss how ringdown measurements can be performed accurately in the presence of frequency fluctuations.
\end{abstract}

\maketitle

\section{Introduction}

Tunable and nonlinear microwave cavities have become ubiquitous in the field of circuit quantum electrodynamics.  Tunable cavities have been used as parametric oscillators to achieve single-shot readout of a superconducting qubit~\cite{Lin2014, Krantz2016}, as quantum buses for selective qubit coupling~\cite{Casparis2019}, as broadband filters \cite{Clark2018}, and as platforms for investigating the dynamical Casimir effect \cite{Wilson2011,Laehteenmaeki2013,Svensson2018}.  Many such cavities are also nonlinear due to the presence of embedded Josephson junctions, but nonlinear microwave cavities on their own have found wide-ranging applications as well.  For example, they have been used as a platform for generating Schrodinger cat states \cite{Kirchmair2013} and as amplifiers operating near the quantum limit of added noise~\cite{Castellanos-Beltran2007,Castellanos-Beltran2008,Bergeal2010}. 

In order to benchmark these devices, however, one must first accurately characterize their basic properties.  Determining their damping rates (or equivalently their quality factors) is particularly important, since these affect the equations of motion governing the cavities such that inaccuracies will have far-reaching consequences.  Unfortunately, a common issue with tunable cavities is that measured damping rates appear to vary with the tuning parameter~\cite{Palacios-Laloy2008, Sandberg2008, Krantz2013, Vissers2015, Kennedy2019, Ferdinand2019}.  Although many have attributed this to causes specific to their devices, here we argue that the tunability itself is at least partially responsible for these observations.  

The line of reasoning is as follows: fluctuations in the tuning parameter will necessarily induce fluctuations in the resonant frequency.  These frequency fluctuations will in turn affect the measurement of scattering matrix elements, which is a standard method for extracting the damping rates of microwave cavities~\cite{Petersan1998, Probst2015}.  A similar line of reasoning holds for cavities with a Kerr nonlinearity: fluctuations in the amplitude of the intra-cavity field will lead to steady-state fluctuations in the resonant frequency, though this effect does not lead to an easily identifiable consequence in the literature as in the case of tunability.

Thus, to accurately determine the damping rates of these cavities from scattering measurements, one must account for the effect of frequency fluctuations.  Here we present a model for this effect, which in the case of tunable cavities both predicts the apparent dependence of the cavity's damping rates on the tuning parameter and allows one to extract the true damping rates in the presence of these frequency fluctuations.  Similarly, in the case of a cavity with a Kerr nonlinearity we analyze the quantum limit of steady-state frequency fluctuations and show that the apparent damping rates can deviate significantly from their true values as the nonlinearity approaches the cavity linewidth.  In both cases we provide bounds on the scale of the fluctuations necessary to yield greater than $10\%$ inaccuracy in the apparent damping rates relative to their true values.  Finally, we analyze how these two sources of fluctuations come together in the case of a tunable cavity with a Kerr nonlinearity.  We expect the use of this model as a fitting function for experimental scattering measurements will allow for more accurate characterization of these cavities moving forward.

In addition to scattering measurements we also consider the effect of frequency fluctuations on ringdown measurements \cite{Reagor2013, Bhupathi2016, thesis_hall, Sinclair2019}, which use the decay of intracavity oscillations to extract the total damping rate of the cavity.  Although they are not sensitive to the individual contributions from multiple loss mechanisms, such measurements are nevertheless useful as an independent metric of damping that could help corroborate the presence of frequency fluctuations by comparison to the results of scattering measurements.  We find that in order to accurately perform a ringdown measurement in the presence of frequency fluctuations, one must directly measure the amplitude or power of the output signal as the cavity rings down, especially if an ensemble average of measurements is necessary to resolve the signal from the noise.  

While our primary goal is the determination of the true damping rates of these cavities, determining the properties of the frequency fluctuations themselves is also important, since these will lead to dephasing \cite{Silveri2017}.  Many recent studies have focused on characterizing the time dependence of frequency fluctuations in microwave cavities \cite{Gao2007, Kumar2008, Barends2008, Barends2009, Lindstroem2011, Burnett2013, Burnett2014, Graaf2014, Burnett2018}, but current methods developed to this end have either assumed that these fluctuations are small compared to the total cavity linewidth or that the damping rates can be determined independently.  Here we show that in the case of tunable and nonlinear microwave cavities both of these assumptions may be broken, and we conclude by extending the most straightforward of these methods to the case of large frequency fluctuations.

\section{Scattering Measurements}

Let $S_{jk}(\Delta)$ be the scattering matrix elements of the microwave network containing the cavity~\cite{text_pozar}, describing an experiment in which the cavity is driven by a sinusoidal signal on port $k$ and measured on port $j$, where $\Delta = \omega - \omega_{0}$ is the detuning of the drive from resonance.  In general, measuring these scattering matrix elements involves a process of averaging, whether implicitly through the time-scale associated with individual measurements or explicitly through the incorporation of multiple independent measurements.  The process of characterizing a cavity then consists of measuring this average value of $S_{jk}(\Delta)$ for a range of detunings around the resonant frequency and comparing these results with a theoretical model to extract the damping rates.

In the presence of fluctuations in the resonant frequency, however, $\omega_{0}\rightarrow \omega_{0} + \delta\omega_{0}(t)$ such that the detuning $\Delta$ may vary over the course of this measurement.  Rather than attempting to take into account the time dependence of $\delta\omega_{0}$ explicitly, we treat these fluctuations as a random variable and model their effect on $S_{jk}(\Delta)$ as an ensemble average.  As a result, in the presence of a fluctuating resonant frequency what will actually be measured is the convolution
\begin{equation}\label{eq:average_scattering_matrix_element}
\overline{S_{jk}}(\Delta) = \int\limits_{-\infty}^{\infty}S_{jk}(\Delta-\Omega)P(\Omega)d\Omega
\end{equation}
where $P(\Omega)$ is the probability density function (PDF) associated with drawing the value $\Omega$ from the random variable $\delta\omega_{0}$.  

There are two generic features of this convolution worth discussing.  First, the average quantity $\overline{S_{jk}}(\Delta)$ will be sensitive to those fluctuations occurring faster than the measurement time and insensitive to those occurring slower than it.  For example, if we can measure $\overline{S_{jk}}(\Delta)$ faster than a given source of frequency noise then our measurement will not be influenced by this noise.  Thus, $P(\Omega)$ should depend on the power spectral density of the frequency fluctuations and the time-scale associated with the measurement.  In particular, since the power spectral density $S_{\omega_{0}\omega_{0}}(f)$ is simply the variance per unit frequency, the variance $\sigma_{\omega_{0}}^{2}$ of frequency fluctuations can be found by integrating over all frequencies $f$.  For a real measurement, however, the low frequency cutoff is set by the inverse of the time $T$ spent measuring $S_{jk}(\Delta)$ at a fixed detuning and the high frequency cutoff is set by the total linewidth of the cavity $\kappa_{\mathrm{tot}}/2\pi$, which determines the maximum rate of cavity response, such that
\begin{equation}\label{eq:variance_PSD_connection}
\sigma_{\omega_{0}}^{2} = \int\limits_{1/T}^{\kappa_{\mathrm{tot}}/2\pi}S_{\omega_{0}\omega_{0}}(f)df.
\end{equation}
We note that it may be possible to extend this simple connection using the Allan variance \cite{Allan1966}, which accounts for dead time between measurements and is well-behaved with respect to the $1/f$ power spectra that are ubiquitous in solid state devices \cite{Paladino2014}, but whose underlying formalism may not directly carry over to the scattering measurements we are considering. 

Second, these scattering matrix elements $S_{jk}(\Delta)$ generally take the form of linear fractional transformations that trace out circles in the complex plane \cite{Marsden1999}.  Since the convolution is essentially a convex combination of points around this circle, the shape traced out by $\overline{S_{jk}}(\Delta)$ will be smaller by comparison, corresponding to apparent changes in the damping rates.  We operationalize the notion of `apparent damping rates' by considering the best fit of the non-fluctuating model $S_{jk}(\Delta)$ to data generated by the fluctuating model $\overline{S_{jk}}(\Delta)$.  

The effect of this convolution can be thought of as a form of inhomogeneous broadening, in that we are effectively averaging over the response functions of an ensemble of cavities with different resonant frequencies.  Similar effects show up in a diverse range of fields, but there are two examples in the same vein as our work worth noting.  First is the Doppler broadening of Lorentzian spectral lines, in which case one obtains a Voigt Profile~\cite{Armstrong1967, Siddons2009}.  Second is the broadening of cavity lineshapes due to a linear coupling with an ensemble of spins that serve as a noisy driving field \cite{Diniz2011}.  Here we apply similar methods to understand the effect of tunability and nonlinearity on the lineshapes of microwave cavities.

\section{Two-sided Cavity}\label{sec:two_sided_cavity}

\begin{figure}
\includegraphics{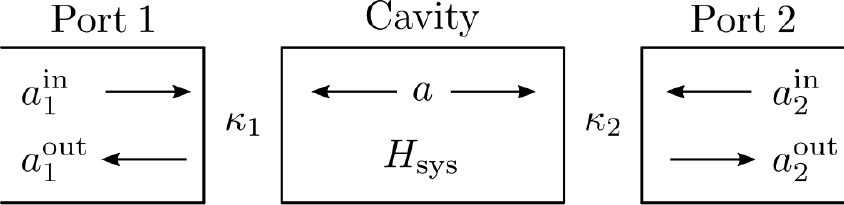}
  \caption{Schematic of a two-sided cavity with Hamiltonian $H_{\mathrm{sys}}$ and damping rates $\kappa_{1}$ and $\kappa_{2}$.}
  \label{fig:two_sided_cavity_schematic}
\end{figure}
As a basis for our study we consider a generic two-sided cavity, depicted schematically in Figure~\ref{fig:two_sided_cavity_schematic}, which we expect to be an appropriate model for a wide variety of systems.  Using input-output theory~\cite{Gardiner1985}, we can relate the fields at the ports of the network to the internal field of the cavity according to
\begin{equation}\label{eq:input_output_relations}
\begin{split}
a_{1}^{\mathrm{out}}(t) - a_{1}^{\mathrm{in}}(t) &= -\sqrt{\kappa_{1}}a(t) \\
a_{2}^{\mathrm{out}}(t) - a_{2}^{\mathrm{in}}(t) &= -\sqrt{\kappa_{2}}a(t)
\end{split}
\end{equation}
where $\kappa_{1}$ and $\kappa_{2}$ are the damping rates associated with their respective ports, such that the internal cavity field obeys the equation of motion
\begin{equation}\label{eq:equation_of_motion}
\begin{split}
\dot{a}(t) = \frac{i}{\hbar}\lbrack H_{\mathrm{sys}}, a \rbrack - \frac{\kappa_{1}+\kappa_{2}}{2} a(t) &+ \sqrt{\kappa_{1}}a_{1}^{\mathrm{in}}(t) \\
&+ \sqrt{\kappa_{2}}a_{2}^{\mathrm{in}}(t).
\end{split}
\end{equation}
We assume for the time being that the cavity is linear, such that $H_{\mathrm{sys}} = \hbar\omega_{0}a^{\dagger}a$ and $\lbrack H_{\mathrm{sys}}, a \rbrack = -\hbar\omega_{0}a$.

In scattering experiments we are generally interested in the steady state response of the cavity as a function of drive frequency $\omega$, to which end we Fourier transform Eq.~\eqref{eq:equation_of_motion} and  solve for the cavity response algebraically
\begin{equation}\label{eq:linear_cavity_response}
\tilde{a}(\omega) = \frac{\sqrt{\kappa_{1}} \tilde{a}_{1}^{\mathrm{in}}(\omega)  + \sqrt{\kappa_{2}} \tilde{a}_{2}^{\mathrm{in}}(\omega)}{i(\omega_{0} - \omega) + (\kappa_{1} + \kappa_{2})/2}
\end{equation}
where a tilde denotes the Fourier transform of the given mode operator.  Plugging this result into the Fourier transform of Eq.~\eqref{eq:input_output_relations}, we find a linear relationship between the input and output mode operators
\begin{equation}
\begin{split}
\tilde{a}_{1}^{\mathrm{out}}(\omega) &= \tilde{a}_{1}^{\mathrm{in}}(\omega) - \frac{\kappa_{1} \tilde{a}_{1}^{\mathrm{in}}(\omega)  + \sqrt{\kappa_{1}\kappa_{2}} \tilde{a}_{2}^{\mathrm{in}}(\omega)}{i(\omega_{0} - \omega) + (\kappa_{1} + \kappa_{2})/2} \\
\tilde{a}_{2}^{\mathrm{out}}(\omega) &= \tilde{a}_{2}^{\mathrm{in}}(\omega) - \frac{\sqrt{\kappa_{1}\kappa_{2}} \tilde{a}_{1}^{\mathrm{in}}(\omega)  + \kappa_{2} \tilde{a}_{2}^{\mathrm{in}}(\omega)}{i(\omega_{0} - \omega) + (\kappa_{1} + \kappa_{2})/2}
\end{split}
\end{equation}
that defines the scattering matrix according to
\begin{equation}
\begin{bmatrix}
\tilde{a}_{1}^{\mathrm{out}}(\omega) \\
\tilde{a}_{2}^{\mathrm{out}}(\omega)
\end{bmatrix}
=
\begin{bmatrix}
S_{11}^{*} & S_{12}^{*} \\
S_{21}^{*} & S_{22}^{*}
\end{bmatrix}
\begin{bmatrix}
\tilde{a}_{1}^{\mathrm{in}}(\omega) \\
\tilde{a}_{2}^{\mathrm{in}}(\omega)
\end{bmatrix}
\end{equation}
where the complex conjugate of each matrix element is taken, by convention, so that its phase corresponds to a counter-clockwise rotation in phase space of the output quadrature operators relative to the inputs.  For ease of notation let $\kappa_{\mathrm{tot}} = \kappa_{1}+\kappa_{2}$ be the total damping rate of the cavity and let $\xi = \kappa_{1}/\kappa_{2}$ be the coupling ratio, in terms of which the scattering matrix elements can be expressed
\begin{equation}
\begin{split}
S_{21}(\Delta) &= -\frac{2\sqrt{\xi}}{1+\xi}\frac{1}{1+2i\Delta/\kappa_{\mathrm{tot}}} \\
S_{12}(\Delta) &= S_{21}(\Delta) \\
S_{11}(\Delta) &= 1 + \sqrt{\xi}S_{21}(\Delta)\\
S_{22}(\Delta) &= 1 + \frac{1}{\sqrt{\xi}}S_{21}(\Delta)
\end{split}
\end{equation}
where $\Delta = \omega - \omega_{0}$ is the detuning, as before.  

As mentioned earlier, our analysis will center around the shape of the trajectories swept out by these scattering matrix elements in the complex plane as a function of detuning, and how these trajectories are deformed by convolving them with a probability distribution, as given by Eq. \eqref{eq:average_scattering_matrix_element}.  To this end, we define the quantities
\begin{equation}
\begin{gathered}
r_{12} = r_{21} = \frac{\sqrt{\xi}}{1+\xi} \\
r_{11} = 1-r_{22} = \frac{\xi}{1+\xi}
\end{gathered}
\end{equation}
and re-express the scattering matrix elements in the form
\begin{equation}\label{eq:ideal_scattering_matrix_elements}
S_{jk}(\Delta ; \kappa_{\mathrm{tot}}, r_{jk}) = \delta_{jk} - r_{jk}\left(1+e^{-2i\arctan(2\Delta/\kappa_{\mathrm{tot}})}\right) 
\end{equation}
where $\delta_{jk}$ is the Kronecker delta. Written in this way, it is clear that each $S_{jk}$ traces out a circular path with radius $r_{jk}$ at a rate determined by $\kappa_{\mathrm{tot}}$.  This is also a convenient expression for analyzing all of the scattering matrix elements at once, since they are equivalent up to translation and a rescaling of $r_{jk}$.  It is worth noting that in order to retrieve $\xi$ from $r_{12}$ or $r_{21}$ one must additionally specify whether $\xi>1$ or $\xi<1$, since they are invariant under $\xi\rightarrow 1/\xi$.  

\section{Tunable Cavity}\label{sec:tunable_cavity}

We now consider a cavity whose resonant frequency $\omega_{0}$ is a function of tuning parameter $x$ such that $\omega_{0} = \omega_{0}(x)$.  In general, this tuning parameter can never be perfectly fixed: it will always fluctuate around its mean value.  To model this effect, let $x \rightarrow \overline{x} + \delta x$, where $\overline{x}$ is the mean value of $x$ and $\delta x$ is a random variable describing fluctuations of $x$ about $\overline{x}$, which we assume to be Gaussian with mean zero and variance $\sigma_{x}^{2}$.  Expanding to lowest order in these fluctuations, we find 
\begin{equation}\label{eq:expanding_frequency_fluctuations}
\omega_{0}(\overline{x}+\delta x) \approx \omega_{0}(\overline{x}) + \delta\omega_{0}(\overline{x})
\end{equation}
where the second term
\begin{equation}\label{eq:gaussian_tuning_fluctuations}
\delta\omega_{0}(\overline{x}) = \frac{\partial\omega_{0}(\overline{x})}{\partial x}\delta x
\end{equation}
is itself a Gaussian random variable with mean zero and variance $\sigma_{\omega_{0}}^{2}$ given by
\begin{equation}\label{eq:tunable_variance}
\sigma_{\omega_{0}}^{2} = \left|\frac{\partial\omega_{0}(\overline{x})}{\partial x}\right|^{2}\sigma_{x}^{2}
\end{equation}
which varies with $\overline{x}$.  The PDF associated with drawing the value $\Omega$ from the random variable $\delta\omega_{0}$ is therefore given by
\begin{equation}\label{eq:gaussian_pdf}
P(\Omega) = \frac{1}{\sqrt{2\pi\sigma_{\omega_{0}}^{2}}}e^{-\Omega^{2}/2\sigma_{\omega_{0}}^{2}}.
\end{equation}

\begin{figure}
\includegraphics{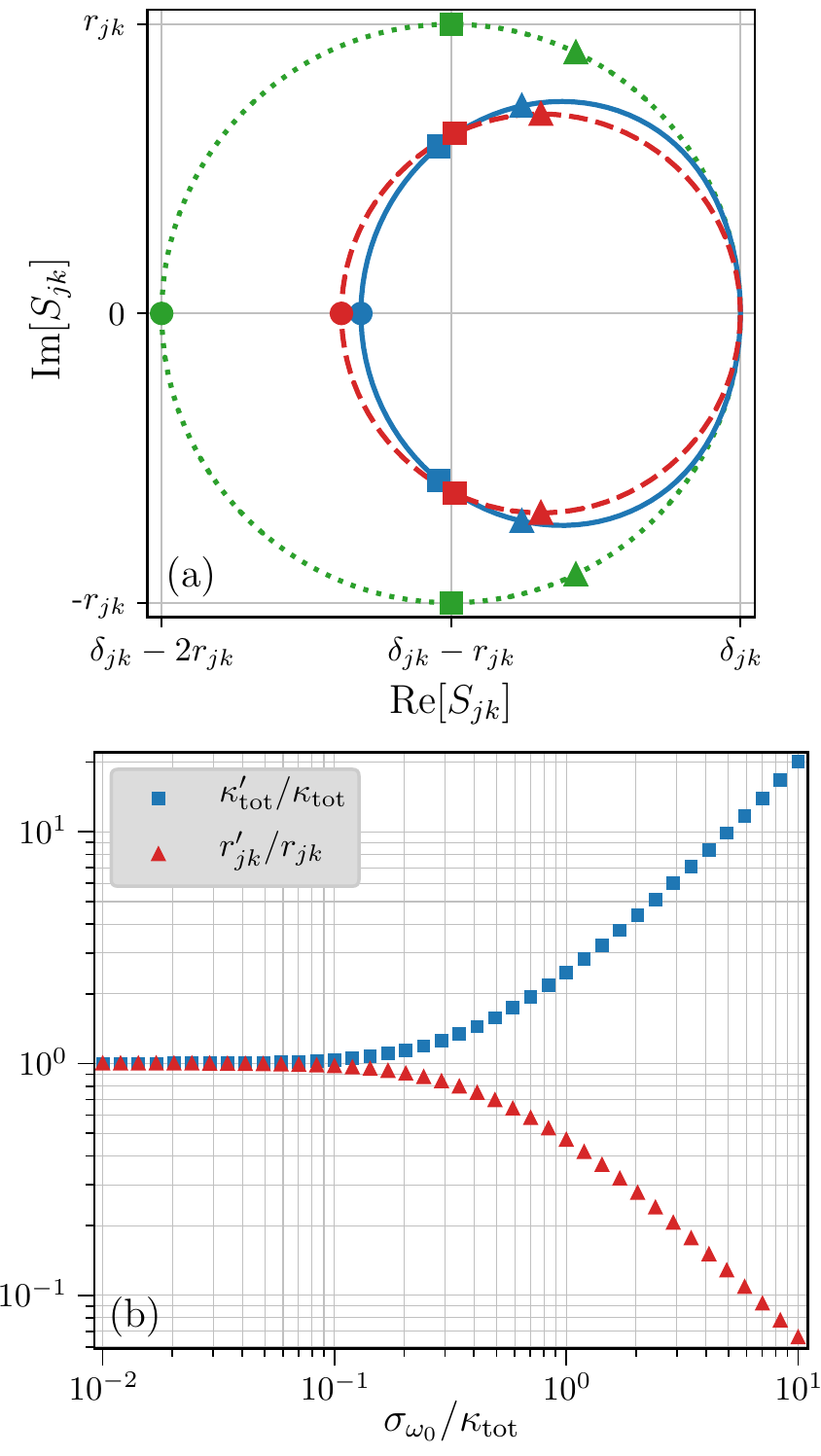}
  \caption{(Color online) The effect of tuning fluctuations expanded to first order on the scattering matrix elements $S_{jk}$.  (a) Visualization of the deformed resonance circle.  Dotted green line: $S_{jk}(\Delta ; \kappa_{\mathrm{tot}}, r_{jk})$.  Solid blue line: $\overline{S_{jk}}(\Delta ; \kappa_{\mathrm{tot}}, r_{jk}, \sigma_{\omega_{0}})$ with $\sigma_{\omega_{0}}=\kappa_{\mathrm{tot}}/2$.  Dashed red line: best fit of $S_{jk}(\Delta ; \kappa_{\mathrm{tot}}^{\prime}, r_{jk}^{\prime})$ to the solid blue trajectory, where $\kappa_{\mathrm{tot}}^{\prime} \approx 1.6\kappa_{\mathrm{tot}}$ and $r_{jk}^{\prime} \approx 0.69 r_{jk}$.  Circles mark $\Delta=0$, squares mark $\Delta = \pm\kappa_{\mathrm{tot}}/2$, and triangles mark $\Delta = \pm\kappa_{\mathrm{tot}}^{\prime}/2$. (b) Results of fitting the model $S_{jk}(\Delta ; \kappa_{\mathrm{tot}}^{\prime}, r_{jk}^{\prime})$ to data generated by $\overline{S_{jk}}(\Delta ; \kappa_{\mathrm{tot}}, r_{jk}, \sigma_{\omega_{0}})$.  The best fit parameters $\kappa_{\mathrm{tot}}^{\prime}$ and $r_{jk}^{\prime}$ quantify the notion of `apparent damping rates.'}
  \label{fig:combined_s11_tuning_linear}
\end{figure}

In the presence of these fluctuations, the average scattering matrix elements are found by plugging Eqs. \eqref{eq:ideal_scattering_matrix_elements} and \eqref{eq:gaussian_pdf} into Eq. \eqref{eq:average_scattering_matrix_element}.  This convolution results in the closed form expression
\begin{multline}\label{eq:avg_sjk_linear_tuning}
\overline{S_{jk}}(\Delta ; \kappa_{\mathrm{tot}}, r_{jk}, \sigma_{\omega_{0}}) = \\
\delta_{jk} - r_{jk}\sqrt{\frac{\pi}{2}}\frac{\kappa_{\mathrm{tot}}}{\sigma_{\omega_{0}}} w\left(\frac{i\kappa_{\mathrm{tot}}-2\Delta}{2\sqrt{2}\sigma_{\omega_{0}}}\right)
\end{multline}
where $w(z)$ is the Faddeeva function, which can be written in terms of the complementary error function as $w(z) = e^{-z^{2}}\mathrm{erfc}(-iz)$~\cite{text_NIST_handbook}.  As illustrated in Fig.~\ref{fig:combined_s11_tuning_linear}(a), the effect of these fluctuations is a deformation of the trajectories of $S_{jk}$ in the complex plane.  Compared to $S_{jk}(\Delta ; \kappa_{\mathrm{tot}}, r_{jk})$, $\overline{S_{jk}}(\Delta ; \kappa_{\mathrm{tot}}, r_{jk}, \sigma_{\omega_{0}})$ is slightly oblong, its apparent radius is smaller, and it traverses its path more slowly as a function of detuning (corresponding to an apparent increase in $\kappa_{\mathrm{tot}}$).  Furthermore, although there is a systematic deviation between the closest fit of $S_{jk}(\Delta ; \kappa_{\mathrm{tot}}^{\prime}, r_{jk}^{\prime})$ to the trajectory generated by $\overline{S_{jk}}(\Delta ; \kappa_{\mathrm{tot}}, r_{jk}, \sigma_{\omega_{0}})$, this deviation is subtle.  It is even more subtle, in fact, if one considers only the squared magnitude of $S_{jk}$ rather than its full complex trajectory as is sometimes done~\cite{Petersan1998, Castellanos-Beltran2007, Sandberg2008}.

Thus, presented with experimental data for which frequency fluctuations are significant, it is quite reasonable to believe it is well-modeled by $S_{jk}(\Delta ; \kappa_{\mathrm{tot}}^{\prime}, r_{jk}^{\prime})$.  If one tries to fit this model to the data, however, one will extract damping rates that appear to vary with $\sigma_{\omega_{0}}$, as illustrated in Fig.~\ref{fig:combined_s11_tuning_linear}(b).  In the case of a tunable cavity for which $\sigma_{\omega_{0}} = |\partial\omega_{0}(\overline{x})/\partial x|\sigma_{x}$, one will therefore find damping rates that seem to vary with $\overline{x}$, which is precisely the syndrome we set out to explain.  By using this model $\overline{S_{jk}}(\Delta ; \kappa_{\mathrm{tot}}, r_{jk}, \sigma_{\omega_{0}})$ as a fitting function for experimental data, however, both the true damping rates and the scale of frequency fluctuations can be determined.  

We emphasize that these fluctuations will always couple into the system as given by Eqs.~\eqref{eq:expanding_frequency_fluctuations}~-~\eqref{eq:tunable_variance}, but they will not always be significant enough to require the use of the fluctuating model $\overline{S_{jk}}(\Delta ; \sigma_{\omega_{0}})$.  The relevant frequency scale for a two-sided cavity is $\kappa_{\mathrm{tot}}$, as seen in Fig.~\ref{fig:combined_s11_tuning_linear}(b): when $\sigma_{\omega_{0}}\ll\kappa_{\mathrm{tot}}$, the apparent and actual damping rates coincide.  As a benchmark in the intermediate case, one must have $\sigma_{\omega_{0}}\lesssim 0.17 \kappa_{\mathrm{tot}}$ in order for both the apparent $\kappa_{\mathrm{tot}}^{\prime}$ and $r_{jk}^{\prime}$ to deviate from their true values $\kappa_{\mathrm{tot}}$ and $r_{jk}$ by less than $10\%$.  

Even when $\sigma_{\omega_{0}}\ll\kappa_{\mathrm{tot}}$, however, frequency fluctuations may still have a non-negligible effect on the apparent damping rates of the cavity.  At maxima and minima of $\omega_{0}(\overline{x})$, for example, $\sigma_{\omega_{0}}$ will vanish and fluctuations in $x$ will only affect the resonant frequency to second order such that
\begin{equation}
\delta\omega_{0}(\overline{x}) = \frac{1}{2}\frac{\partial^{2}\omega_{0}(\overline{x})}{\partial x^{2}}\delta x^{2}.
\end{equation}
Making the same assumption that $\delta x$ is Gaussian distributed with mean zero and variance $\sigma_{x}^{2}$, these fluctuations will follow a chi squared distribution with one degree of freedom.  If we define the scale associated with these fluctuations to be
\begin{equation}
D = \frac{\sigma_{x}^{2}}{2}\frac{\partial^{2}\omega_{0}(\overline{x})}{\partial x^{2}}
\end{equation}
then the probability distribution associated with this random variable can be expressed as
\begin{equation}
P(\Omega) = \frac{1}{\sqrt{2\pi D\Omega}}e^{-\Omega/2D} \Theta(\Omega/D)
\end{equation}
where $\Theta$ is the Heaviside step function, such that $P(\Omega)$ has support on either the positive or negative real axis depending on the sign of $D$.  Performing the convolution given by Eq. \eqref{eq:average_scattering_matrix_element} for this case, we find
\begin{multline}
\overline{S_{jk}}(\Delta ; \kappa_{\mathrm{tot}}, r_{jk}, D) = \\
\delta_{jk} - i r_{jk} \frac{\sqrt{\pi}}{2}\frac{\kappa_{\mathrm{tot}}}{D}\frac{w\left(i\sqrt{(i\kappa_{\mathrm{tot}}-2\Delta)/4D}\right)}{\sqrt{(i\kappa_{\mathrm{tot}}-2\Delta)/4D}}
\end{multline}
where the branch cut can be made on the negative real axis without any issue since $\kappa_{\mathrm{tot}} > 0$.  

\begin{figure}
\includegraphics{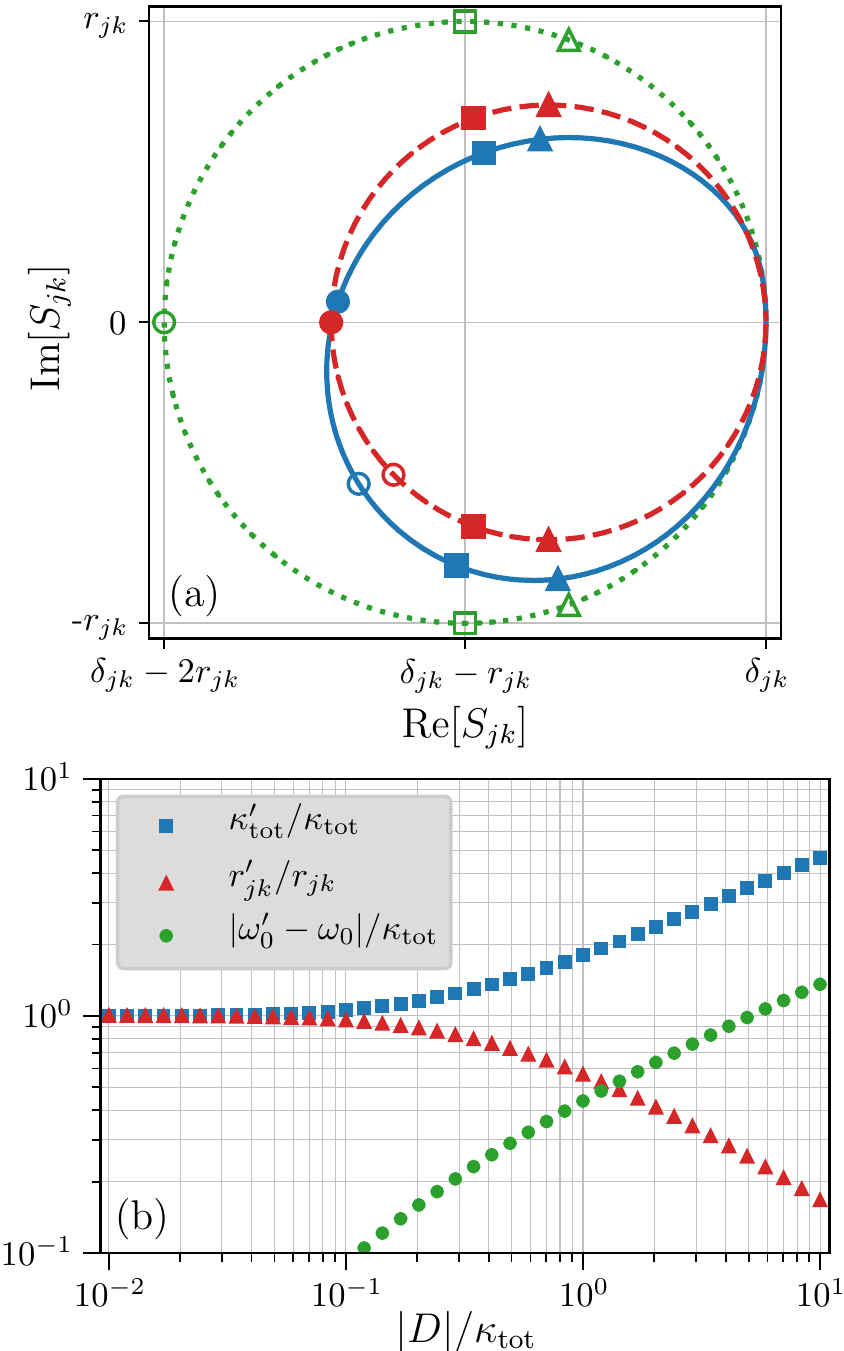}
  \caption{(Color online) The effect of tuning fluctuations expanded to second order on the scattering matrix elements $S_{jk}$.  (a) Visualization of the deformed resonance circle.  Dotted green line: $S_{jk}(\Delta ; \kappa_{\mathrm{tot}}, r_{jk})$.  Solid blue line: $\overline{S_{jk}}(\Delta ; \kappa_{\mathrm{tot}}, r_{jk}, D)$ with $D=\kappa_{\mathrm{tot}}/2$.  Dashed red line: best fit of $S_{jk}(\Delta^{\prime} ; \kappa_{\mathrm{tot}}^{\prime}, r_{jk}^{\prime})$ to the solid blue trajectory where $\omega_{0}^{\prime}-\omega_{0}\approx 0.29\kappa_{\mathrm{tot}}$, $\kappa_{\mathrm{tot}}^{\prime} \approx 1.4\kappa_{\mathrm{tot}}$, and $r_{jk}^{\prime} \approx 0.72 r_{jk}$.  Hollow circles mark $\Delta=0$, hollow squares mark $\Delta = \pm\kappa_{\mathrm{tot}}/2$, and hollow triangles mark $\Delta = \pm\kappa_{\mathrm{tot}}^{\prime}/2$.  Filled shapes mark the corresponding points for $\Delta^{\prime}$. (b) Results of fitting the model $S_{jk}(\Delta^{\prime} ; \kappa_{\mathrm{tot}}^{\prime}, r_{jk}^{\prime})$ to data generated by $\overline{S_{jk}}(\Delta ; \kappa_{\mathrm{tot}}, r_{jk}, D)$.  Values of $|\omega_{0}^{\prime}-\omega_{0}|<\kappa_{\mathrm{tot}}/10$ have been omitted for clarity.}
  \label{fig:combined_s11_tuning_quadratic}
\end{figure}

As can be seen in Fig. \ref{fig:combined_s11_tuning_quadratic}(a), the effect of these fluctuations is a deformation of $S_{jk}$ that once again leads to an increase in the apparent total damping rate $\kappa_{\mathrm{tot}}$ and a decrease in the apparent radius $r_{jk}$.  Unlike the case of Gaussian fluctuations, however, $\overline{S_{jk}}(\Delta ; \kappa_{\mathrm{tot}}, r_{jk}, D)$ is now asymmetric with respect to the real axis as a result of the probability distribution only having support over the positive or negative reals, which also gives rise to an apparent shift in the resonant frequency.  We have only displayed the case of $D>0$ for simplicity; the corresponding trajectory for $D<0$ can be visualized using the relationship
\begin{equation}\label{eq:reflect_trajectory}
\overline{S_{jk}}(\Delta ; \kappa_{\mathrm{tot}}, r_{jk}, -D) = \overline{S_{jk}}(-\Delta ; \kappa_{\mathrm{tot}}, r_{jk}, D)^{*}
\end{equation}
which amounts to reflecting the shape of the trajectory across the real axis and traversing it in the same clockwise orientation.  As before, we obtain the apparent cavity parameters as a function of $D$ by fitting the model $S_{jk}(\Delta^{\prime} ; \kappa_{\mathrm{tot}}^{\prime}, r_{jk}^{\prime})$ to data generated by $\overline{S_{jk}}(\Delta ; \kappa_{\mathrm{tot}}, r_{jk}, D)$, the results of which are displayed in Fig. \ref{fig:combined_s11_tuning_quadratic}(b).  In this case, we find that in order for the apparent $\kappa_{\mathrm{tot}}^{\prime}$ and $r_{jk}^{\prime}$ to both be within $10\%$ of their true values $\kappa_{\mathrm{tot}}$ and $r_{jk}$, we must have $|D|\lesssim 0.14\kappa_{\mathrm{tot}}$.

Thus, if we tune the cavity to a point $\overline{x}_{0}$ where $\sigma_{\omega_{0}}(\overline{x}_{0})=0$ and $|D(\overline{x}_{0})|\ll\kappa_{\mathrm{tot}}$, the effect of frequency fluctuations will be negligible and we can extract the correct damping rates of the cavity using the non-fluctuating model $S_{jk}(\Delta)$.  In certain cases it may be sufficient to characterize the cavity only at such points, but in many cases it may be either preferable or necessary to characterize the system over a wide range of the tuning parameter $\overline{x}$.  For example, if the cavity is tunable over a range of frequencies far greater than its linewidth then the physical properties of both the system and its environment (such as characteristic impedances) are likely to vary appreciably with the operating frequency $\omega_{0}(\overline{x})$, leading to frequency-dependent damping rates.  This model for $\overline{S_{jk}}$ allows one to extract this dependence without it being obscured by the presence of frequency fluctuations.

\section{Kerr Cavity}\label{sec:kerr_cavity}

We next consider a cavity with a Kerr nonlinearity $K$, such that the system Hamiltonian takes the form
\begin{equation}
H_{\mathrm{sys}} = \hbar\omega_{0}a^{\dagger}a + \frac{\hbar K}{2}a^{\dagger 2}a^{2}.
\end{equation}
The quantum Langevin equation \eqref{eq:equation_of_motion} for the operator $a$ then becomes
\begin{equation}
\begin{split}
\dot{a}(t) =& -i\left[\omega_{0} + K a^{\dagger}(t)a(t) \right]a(t) \\
&- \frac{\kappa_{1} + \kappa_{2}}{2}a(t) + \sqrt{\kappa_{1}}a_{1}^{\mathrm{in}}(t) + \sqrt{\kappa_{2}}a_{2}^{\mathrm{in}}(t)
\end{split}
\end{equation}
from which we see that the resonant frequency of the cavity depends on the state of the cavity according to 
\begin{equation}
\omega_{0}(a,a^{\dagger}) = \omega_{0} + K a^{\dagger}a.
\end{equation}
We will restrict our focus to the regime $|K|<\kappa_{\mathrm{tot}}$ so a semi-classical treatment is appropriate \cite{Yurke2006}, and further assume that $kT\ll\hbar\omega_{0}$ such that the steady state response of the cavity is a coherent state to good approximation (rather than a displaced thermal state).   We also assume that characterization of this system will be performed with a sufficiently weak driving field such that $|K|\langle a^{\dagger}a\rangle \ll \kappa_{\mathrm{tot}}$ in the steady state \cite{Krantz2013}.  In this case the cavity response will be linear to a good approximation and the scattering matrix elements (in the absence of frequency fluctuations) will be given by Eq. \eqref{eq:ideal_scattering_matrix_elements}.  

Since the scattered signal carries information about the quadratures $X_{1} = a^{\dagger}+a$ and $X_{2} = i(a^{\dagger}-a)$ of the intracavity field \cite{Clerk2010, Silva2010, Eichler2012}, the fluctuations in these operators are what will affect our measurement of the scattering matrix elements.  We therefore express the resonant frequency as
\begin{equation}
\omega_{0}(X_{1}, X_{2}) = \omega_{0} + \frac{K}{4}(X_{1}^{2} + X_{2}^{2})
\end{equation}
where we've absorbed an overall constant into $\omega_{0}$.  Fluctuations $\delta X_{1,2} = X_{1,2} - \langle X_{1,2} \rangle$ of these operators in the steady state will be independent and Gaussian, each with zero mean and unit variance.  The resulting frequency fluctuations $\delta\omega_{0} = \omega_{0}(X_{1}, X_{2}) - \omega_{0}$ will follow a non-central chi squared distribution with two degrees of freedom, whose PDF is given by 
\begin{equation}\label{eq:noncentral_chisq_pdf}
P(\Omega) = \frac{2}{|K|}e^{-2\left(n + \Omega/K\right)} I_{0}\left(4\sqrt{n\Omega/K}\right)\Theta(\Omega/K)
\end{equation}
where $n = \langle a^{\dagger}a \rangle = \left(\langle X_{1} \rangle^{2} + \langle X_{2} \rangle^{2}\right)/4$ is the average number of photons in the cavity (assuming a coherent steady state) and $I_{0}(z)$ is the zeroth modified Bessel function of the first kind.  Note that the number of photons $n$ is related to the non-centrality parameter $\lambda$ of the non-central chi squared distribution according to $\lambda = 4n$.  There are two key downsides to working with this probability distribution.  First, the convolution given by Eq. \eqref{eq:average_scattering_matrix_element} does not readily simplify into an expression in terms of special functions (for which efficient implementations exist in most programming languages), so the numerical integration must be implemented manually.  Second, if we try to fit this model (given by Eqs. \eqref{eq:average_scattering_matrix_element} and \eqref{eq:noncentral_chisq_pdf}) to experimental data for the scattering matrix element $S_{jk}$, then we would have two parameters ($K$ and $n$) that govern the subtle deviation of the fluctuating model $\overline{S_{jk}}(\Delta)$ from the non-fluctuating model rather than one.  This may lead to overfitting problems unless either $K$ or $n$ can be determined independently.  

\begin{figure}
\includegraphics{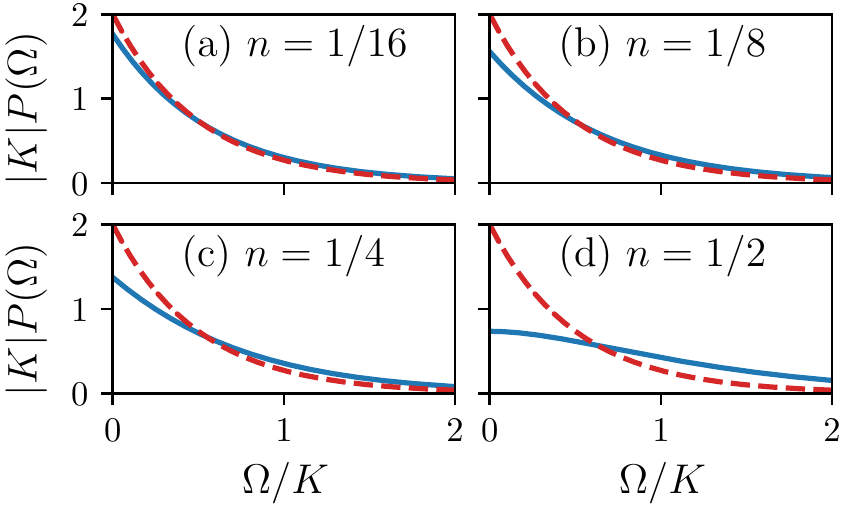}
  \caption{(Color online) Approximating the PDF of a non-central chi squared distribution as a central chi squared distribution (both with two degrees of freedom) for increasing intracavity photon occupation $n = \langle a^{\dagger} a \rangle$.  Solid blue lines are the exact PDF given by Eq. \eqref{eq:noncentral_chisq_pdf}, dashed red lines are the approximate PDF given by Eq. \eqref{eq:chisq_pdf}.}
  \label{fig:kerr_noncentral_chisq_approximation}
\end{figure}

It is therefore convenient to additionally work in a regime where $n \ll 1$, such that fluctuations in the resonant frequency can be approximated as
\begin{equation}\label{eq:chisq_kerr_fluctuations}
\delta\omega_{0} \approx \frac{K}{4}\left(\delta X_{1}^{2} + \delta X_{2}^{2}\right).
\end{equation}
which will follow a central chi-squared distribution with two degrees of freedom.  Since the variance of the non-central chi squared distribution increases with $n$ and we've already taken the limit of $kT/\hbar\omega_{0}\rightarrow 0$, this limit as $n\rightarrow 0$ can be thought of as the quantum limit of steady-state frequency fluctuations in a Kerr cavity.  The probability of drawing the value $\Omega$ from the random variable $\delta\omega_{0}$ in this case is given by
\begin{equation}\label{eq:chisq_pdf}
P(\Omega) = \frac{2}{|K|}e^{-2\Omega/K}\:\Theta(\Omega/K)
\end{equation}
where $\Theta$ is the Heaviside step function, such that $P(\Omega)$ has support on either the positive or negative real axis depending on the sign of $K$.  As seen in Fig. \ref{fig:kerr_noncentral_chisq_approximation}, this approximation works very well up to $n \sim 1/8$ and accumulates significant errors by $n \sim 1/2$.  For this probability distribution, Eq. \eqref{eq:average_scattering_matrix_element} simplifies to 
\begin{multline}\label{eq:avg_sjk_kerr}
\overline{S_{jk}}(\Delta ; \kappa_{\mathrm{tot}}, r_{jk}, K) = \\
\delta_{jk} - 2ir_{jk}\frac{\kappa_{\mathrm{tot}}}{K}e^{(i\kappa_{\mathrm{tot}}-2\Delta)/K}\Gamma\Bigl(0,\frac{i\kappa_{\mathrm{tot}}-2\Delta}{K}\Bigr)
\end{multline}
where $\Gamma(a, z) = \int_{z}^{\infty}t^{a-1}e^{-t}dt$ is the incomplete gamma function \cite{text_NIST_handbook}.

\begin{figure}
\includegraphics{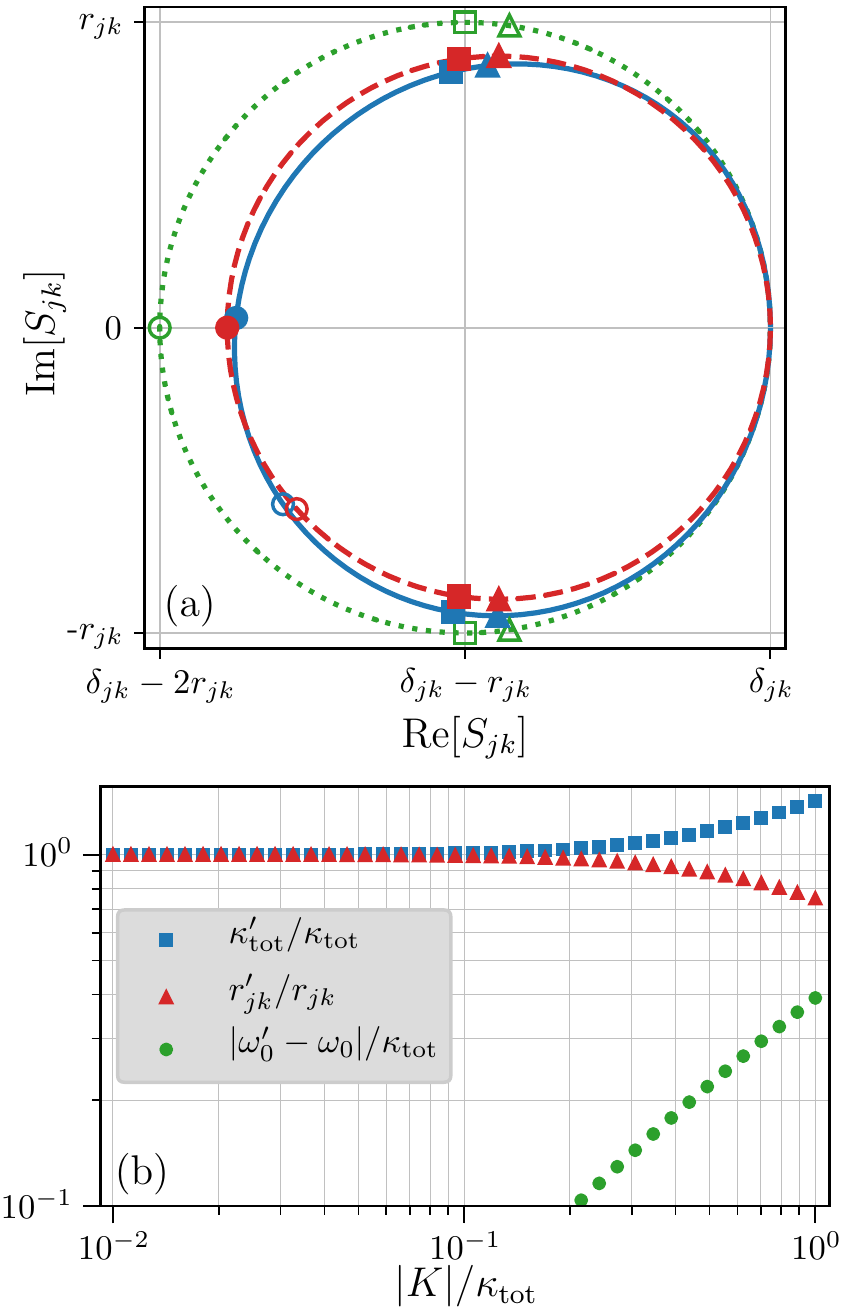}
  \caption{(Color online) The effect of quantum fluctuations on the scattering matrix elements $S_{jk}$ of a Kerr cavity when $n = \langle a^{\dagger}a \rangle \ll 1$.  (a) Visualization of the deformed resonance circle.  Dotted green line: $S_{jk}(\Delta ; \kappa_{\mathrm{tot}}, r_{jk})$.  Solid blue line: $\overline{S_{jk}}(\Delta ; \kappa_{\mathrm{tot}}, r_{jk}, K)$ with $K=\kappa_{\mathrm{tot}}/2$.  Dashed red line: best fit of $S_{jk}(\Delta^{\prime} ; \kappa_{\mathrm{tot}}^{\prime}, r_{jk}^{\prime})$ to the solid blue trajectory where $\omega_{0}^{\prime} - \omega_{0} \approx 0.22\kappa_{\mathrm{tot}}$, $\kappa_{\mathrm{tot}}^{\prime} \approx 1.2\kappa_{\mathrm{tot}}$, and $r_{jk}^{\prime} \approx 0.89 r_{jk}$.  Hollow circles mark $\Delta=0$, hollow squares mark $\Delta = \pm\kappa_{\mathrm{tot}}/2$, and hollow triangles mark $\Delta = \pm\kappa_{\mathrm{tot}}^{\prime}/2$.  Filled shapes mark the corresponding points for $\Delta^{\prime}$. (b) Results of fitting the model $S_{jk}(\Delta^{\prime} ; \kappa_{\mathrm{tot}}^{\prime}, r_{jk}^{\prime})$ to data generated by $\overline{S_{jk}}(\Delta ; \kappa_{\mathrm{tot}}, r_{jk}, K)$.  Values of $|\omega_{0}^{\prime}-\omega_{0}|<\kappa_{\mathrm{tot}}/10$ have been omitted for clarity.}
  \label{fig:combined_s11_kerr}
\end{figure}

As before, the effect of these fluctuations is an apparent increase in the total damping rate $\kappa_{\mathrm{tot}}$ and decrease in the radius $r_{jk}$ of the resonance circle, as seen in Fig. \ref{fig:combined_s11_kerr}(a).  Suprisingly, even as $K$ approaches the cavity linewidth $\kappa_{\mathrm{tot}}$ the deviation of the resulting scattering matrix elements $\overline{S_{jk}}$ from the non-fluctuating model $S_{jk}$ remains subtle.  As such, this systematic deviation is easily obscured by measurement noise and is therefore likely to be overlooked.  Just as in the case of a chi squared distribution with one degree of freedom $P(\Omega)$ only has support on either the positive or negative reals, which leads to both a shift in the apparent resonant frequency and asymmetry in the shape of the resonance circle, but these effects are slight.  The case of $K>0$ is displayed for simplicity, but the corresponding resonance circle for $K<0$ can again be visualized by using Eq. \eqref{eq:reflect_trajectory} with the substitution $D\rightarrow K$.  By fitting the model $S_{jk}(\Delta^{\prime} ; \kappa_{\mathrm{tot}}^{\prime}, r_{jk}^{\prime})$ to data generated by $\overline{S_{jk}}(\Delta ; \kappa_{\mathrm{tot}}, r_{jk}, K)$ we have obtained the apparent values of $\kappa_{\mathrm{tot}}^{\prime}$ and $r_{jk}^{\prime}$ as a function of $|K|$, which we present in Fig. \ref{fig:combined_s11_kerr}(b).  We find that in order for the apparent $\kappa_{\mathrm{tot}}^{\prime}$ and $r_{jk}^{\prime}$ to both be within $10\%$ of their true values $\kappa_{\mathrm{tot}}$ and $r_{jk}$, we must have $|K|\lesssim 0.35\kappa_{\mathrm{tot}}$.  Thus, this effect only becomes significant as $|K|$ approaches the cavity linewidth.  

Another important case to consider is that of a cavity with a modest Kerr nonlinearity $K$ driven at high enough powers such that the number of photons $n$ in the cavity is non-negligible, a situation that is particularly important for experiments in which it is either necessary or useful to characterize the cavity using a range of input powers.  In this case there is not a closed form expression for the average scattering matrix elements; instead, they must be obtained by numerically evaluating the convolution
\begin{multline}\label{eq:avg_sjk_noncentral_chisq}
\overline{S_{jk}}(\Delta ; \kappa_{\mathrm{tot}}, r_{jk}, K, n) = \\
\int\limits_{-\infty}^{\infty}S_{jk}(\Delta-\Omega ; \kappa_{\mathrm{tot}}, r_{jk}) P(\Omega ; K, n)d\Omega
\end{multline}
where $S_{jk}(\Delta-\Omega ; \kappa_{\mathrm{tot}}, r_{jk})$ is given by Eq. \eqref{eq:ideal_scattering_matrix_elements} and $P(\Omega ; K, n)$ is given by Eq. \eqref{eq:noncentral_chisq_pdf}.  We analyze the deformation induced by this convolution for fixed Kerr nonlinearity $K_{0} = \kappa_{\mathrm{tot}}/10$, for which the deformation is negligible in the limit $n\rightarrow 0$ (see Fig. \ref{fig:combined_s11_kerr}(b)).

\begin{figure}
\includegraphics{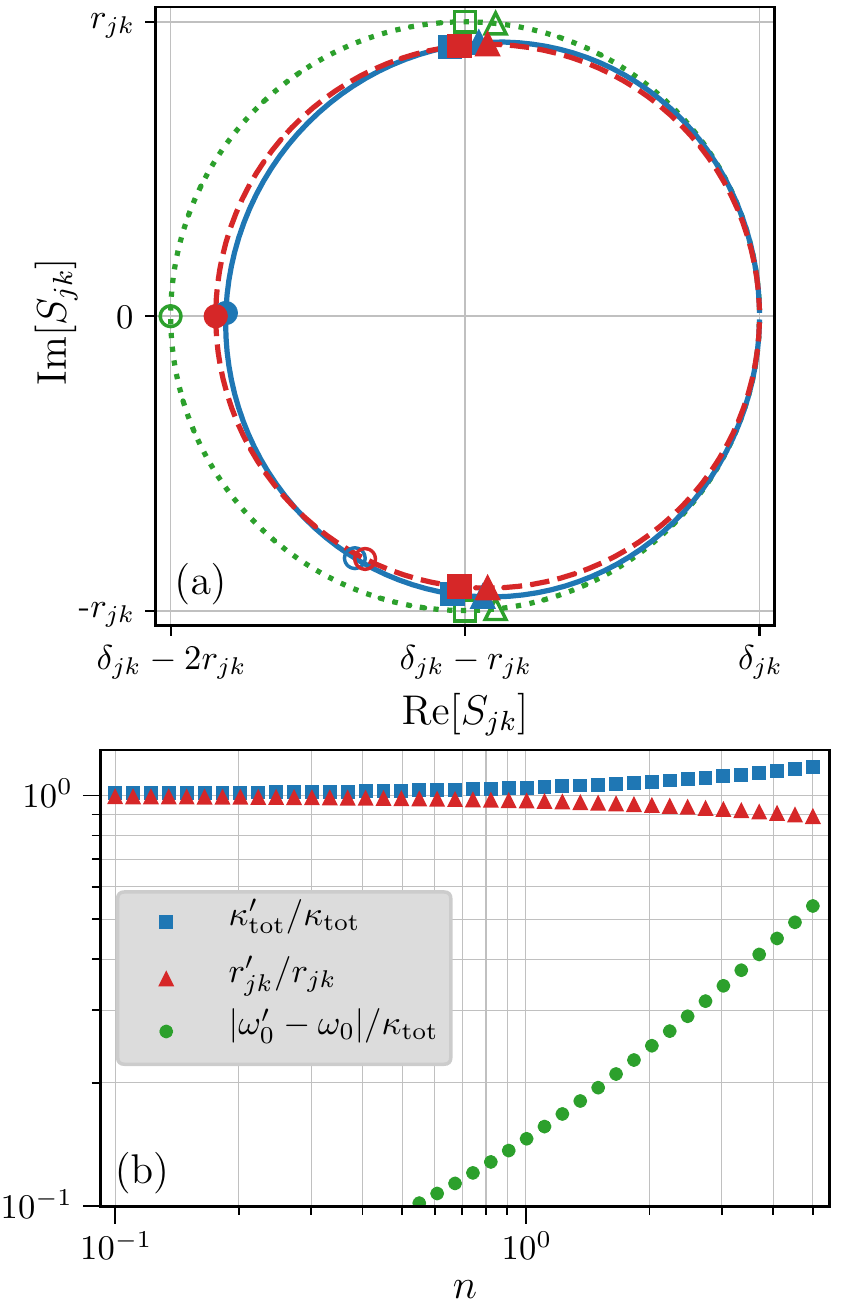}
  \caption{(Color online) The effect of quantum fluctuations on the scattering matrix elements $S_{jk}$ of a cavity with fixed Kerr nonlinearity $K_{0} = \kappa_{\mathrm{tot}}/10$ and non-negligible cavity occupation $n = \langle a^{\dagger}a \rangle$.  (a) Visualization of the deformed resonance circle.  Dotted green line: $S_{jk}(\Delta ; \kappa_{\mathrm{tot}}, r_{jk})$.  Solid blue line: $\overline{S_{jk}}(\Delta ; \kappa_{\mathrm{tot}}, r_{jk}, K_{0}, n)$ with $n=3$ photons.  Dashed red line: best fit of $S_{jk}(\Delta^{\prime} ; \kappa_{\mathrm{tot}}^{\prime}, r_{jk}^{\prime})$ to the solid blue trajectory where $\omega_{0}^{\prime} - \omega_{0} \approx 0.34\kappa_{\mathrm{tot}}$, $\kappa_{\mathrm{tot}}^{\prime} \approx 1.1\kappa_{\mathrm{tot}}$, and $r_{jk}^{\prime} \approx 0.92 r_{jk}$.  Hollow circles mark $\Delta=0$, hollow squares mark $\Delta = \pm\kappa_{\mathrm{tot}}/2$, and hollow triangles mark $\Delta = \pm\kappa_{\mathrm{tot}}^{\prime}/2$.  Filled shapes mark the corresponding points for $\Delta^{\prime}$. (b) Results of fitting the model $S_{jk}(\Delta^{\prime} ; \kappa_{\mathrm{tot}}^{\prime}, r_{jk}^{\prime})$ to data generated by $\overline{S_{jk}}(\Delta ; \kappa_{\mathrm{tot}}, r_{jk}, K_{0}, n)$ as a function of cavity photons $n$.  Values of $|\omega_{0}^{\prime}-\omega_{0}|<\kappa_{\mathrm{tot}}/10$ have been omitted for clarity.}
  \label{fig:combined_s11_noncentral_chisq}
\end{figure}

As with the previous cases, this convolution yields an apparent increase in the total damping rate $\kappa_{\mathrm{tot}}$ and decrease in the radius $r_{jk}$ of the resonance circle, as seen in Fig. \ref{fig:combined_s11_noncentral_chisq}(a).  Furthermore, the shape of the deformed scattering matrix element $\overline{S_{jk}}$ is qualitatively similar to that in the case of a stronger Kerr nonlinearity $K$ and negligible cavity occupation $n$ (compare with Fig. \ref{fig:combined_s11_kerr}(a)).  And, once again, the resulting deviation of the scattering matrix elements $\overline{S_{jk}}$ from the non-fluctuating model $S_{jk}$ is subtle to the point of being easily overlooked or obscured by measurement noise.  By fitting the model $S_{jk}(\Delta^{\prime} ; \kappa_{\mathrm{tot}}^{\prime}, r_{jk}^{\prime})$ to data generated by $\overline{S_{jk}}(\Delta ; \kappa_{\mathrm{tot}}, r_{jk}, K_{0}, n)$ we have obtained the apparent values of $\kappa_{\mathrm{tot}}^{\prime}$ and $r_{jk}^{\prime}$ as a function of cavity occupation $n = \langle a^{\dagger}a \rangle$, which we present in Fig. \ref{fig:combined_s11_noncentral_chisq}(b).  We find that in order for the apparent $\kappa_{\mathrm{tot}}^{\prime}$ and $r_{jk}^{\prime}$ to both be within $10\%$ of their true values $\kappa_{\mathrm{tot}}$ and $r_{jk}$ when $K = \kappa_{\mathrm{tot}}/10$, we must have $n\lesssim 2.6$.  It is important to note that this effect is not scale-free; it will become stronger (weaker) for increasing (decreasing) values of $K/\kappa_{\mathrm{tot}}$, even if $nK/\kappa_{\mathrm{tot}}$ is held constant.  

Thus, for cavities with even modest Kerr nonlinearities $K\gtrsim \kappa_{\mathrm{tot}}/10$ operated at the few-photon level, one will extract damping rates that differ appreciably from their true values if the effect of quantum fluctuations on the scattering matrix elements is not properly taken into account.  Numerous examples of Kerr cavities in this regime are described in the literature \cite{Hoffman2011, Zakka-Bajjani2011, Krantz2013, Kirchmair2013, Svensson2017, Bengtsson2018, Wang2019}; based on our work, we believe that there are likely discrepancies between the reported damping rates of such cavities and their true values.  The fact that these apparent damping rates vary with the cavity occupation $n$ is particularly important to account for experimentally since it is well known that the true damping rates of microwave cavities vary with $n$ as well, due to the two-level systems present in the dielectric substrates on which planar cavities are fabricated \cite{Martinis2005, Barends2010, Vissers2010, Sage2011, Khalil2011, Megrant2012, Bruno2015}.  As we've shown, however, these discrepancies will persist even in the limit of zero cavity occupation.  By using the model presented in this section, these systematic errors can be mitigated.

\section{Tunable Kerr Cavity}\label{sec:tunable_kerr_cavity}

It is useful to consider the combined effect of tunability and the Kerr nonlinearity on frequency fluctuations, since many tunable cavities contain Josephson junctions which introduce nonlinearity into the system.  Making the same assumptions as in Sections \ref{sec:tunable_cavity} and \ref{sec:kerr_cavity}, which led to Eqs. \eqref{eq:gaussian_tuning_fluctuations} and \eqref{eq:chisq_kerr_fluctuations}, the frequency fluctuations in this case take the form
\begin{equation}
\delta\omega_{0} = \frac{\partial\omega_{0}(\overline{x})}{\partial x}\delta x + \frac{K}{4}\left(\delta X_{1}^{2} + \delta X_{2}^{2}\right)
\end{equation}
which is the sum of a Gaussian-distributed and a chi squared-distributed random variable.  The resulting PDF is the convolution of a Gaussian and chi-squared distribution
\begin{equation}\label{eq:tunable_kerr_pdf_exact}
\begin{split}
P(\Omega) &= \int\limits_{-\infty}^{\infty}\frac{2e^{-2\Omega^{\prime}/K}}{|K|}\frac{e^{-(\Omega-\Omega^{\prime})^{2}/2\sigma_{\omega_{0}}^{2}}}{\sqrt{2\pi\sigma_{\omega_{0}}^{2}}}\Theta(\Omega^{\prime}/K)d\Omega^{\prime} \\
&= e^{2\sigma_{\omega_{0}}^{2}/K^{2} - 2\Omega/K}\left[\frac{1}{|K|} - \frac{\mathrm{erf}\left(\frac{\sqrt{2}\sigma_{\omega_{0}}}{K} - \frac{\Omega}{\sqrt{2}\sigma_{\omega_{0}}}\right)}{K}\right]
\end{split}
\end{equation}
where $\sigma_{\omega_{0}}$ is again given by Eq. \eqref{eq:tunable_variance}.  Working with this probability distribution has the same downsides as the non-central chi squared distribution: the convolution of Eq. \eqref{eq:average_scattering_matrix_element} does not readily simplify in terms of special functions, and it depends on two extra free parameters rather than one.  When $|K|/\sigma_{\omega_{0}}$ is either very small or very large, however, we can once again obtain simple expressions for the average scattering matrix elements.

\begin{figure}
\includegraphics{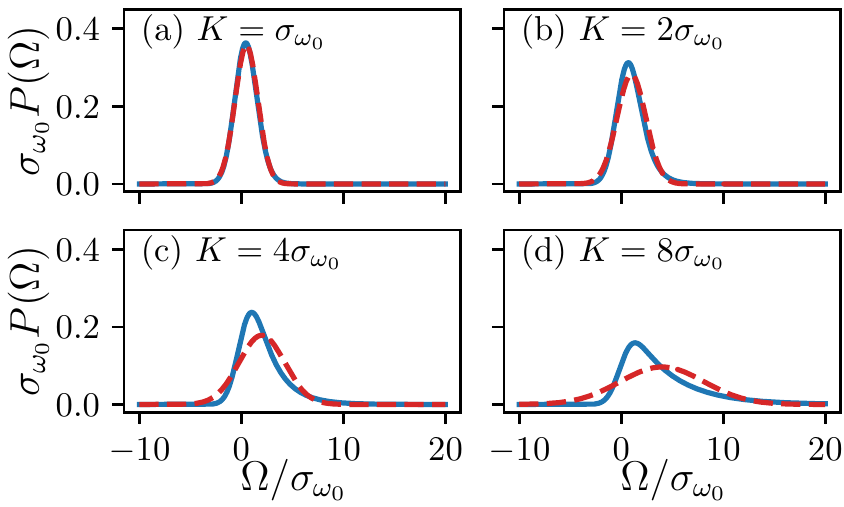}
  \caption{(Color online) Approximating the PDF of the tunable Kerr cavity with a Gaussian for increasing values of $K$.  Solid blue lines are the exact PDF given by Eq. \eqref{eq:tunable_kerr_pdf_exact}, dashed red lines are the approximate Gaussian PDF given by Eq. \eqref{eq:tunable_kerr_pdf_approximate}.}
  \label{fig:tunable_kerr_gaussian_approximation}
\end{figure}

When $|K| \ll \sigma_{\omega_{0}}$, the resulting probability distribution will be well-approximated by a Gaussian.  This is most easily seen by expanding this distribution's cumulant generating function to second order in $K$
\begin{equation}
\begin{split}
C(t) &= \frac{1}{2}\sigma_{\omega_{0}}^{2}t^{2} - \log\left(1-\frac{Kt}{2}\right) \\
&\approx \frac{1}{2}Kt + \frac{1}{2}\left(\sigma_{\omega_{0}}^{2} + \frac{K^{2}}{4}\right)t^{2}
\end{split}
\end{equation}
which is identical to that of a Gaussian with mean $K/2$ and variance $\sigma_{\omega_{0}}^{2} + K^{2}/4$.  More precisely, we can expand Eq. \eqref{eq:tunable_kerr_pdf_exact} as an Edgeworth series \cite{Wallace1958} such that
\begin{equation}\label{eq:tunable_kerr_pdf_approximate}
P(\Omega) \approx \frac{\exp\left[\cfrac{-(\Omega - K/2)^{2}}{2(\sigma_{\omega_{0}}^{2} + K^{2}/4)}\right]}{\sqrt{2\pi(\sigma_{\omega_{0}}^{2} + K^{2}/4)}}
\end{equation}
to lowest order.  Since the higher order cumulants will be $c_{m} = (K/2)^{m}$ for integers $m>3$, the leading order correction to this expression will go as $K^{3}/(4\sigma_{\omega_{0}}^{2}+K^{2})^{3/2}$.  Although this approximation is most accurate when $|K| \ll \sigma_{\omega_{0}}$, it actually works very well up to $|K| \approx 2\sigma_{\omega_{0}}$ and only begins to significantly deviate from the exact result when $|K| \gtrsim 4\sigma_{\omega_{0}}$, as can be seen in Fig. \ref{fig:tunable_kerr_gaussian_approximation}.  Thus, for $|K| \lesssim 2\sigma_{\omega_{0}}$, the resonant frequency of our cavity will be shifted by $K/2$ and the average scattering matrix elements can be approximated by Eq. \eqref{eq:avg_sjk_linear_tuning} with $\sigma_{\omega_{0}} \rightarrow \sqrt{\sigma_{\omega_{0}}^{2} + K^{2}/4}$.  In the opposite regime, when $|K|\gg\sigma_{\omega_{0}}$, the cumulants of the chi squared distribution will dominate at all orders and the scattering matrix elements can be approximated by Eq. \eqref{eq:avg_sjk_kerr} if $n \ll 1$, and by Eq. \eqref{eq:avg_sjk_noncentral_chisq} otherwise.

\section{Ringdown Measurements}

We now consider the effect of frequency fluctuations on ringdown measurements \cite{Reagor2013, Bhupathi2016, thesis_hall, Sinclair2019}.  This type of measurement is performed by driving the cavity on resonance to a steady state amplitude, turning off the drive, and observing the subsequent decay of the intracavity field, from which the total damping rate of the cavity can be determined.  Before the measurement, the average resonant frequency $\overline{\omega_{0}}$ must first be measured, which sets the drive frequency, after which the actual resonant frequency $\omega_{0}(t)$ may fluctuate from this average value by $\delta\omega_{0}(t) = \omega_{0}(t) - \overline{\omega_{0}}$ during the ringdown measurement.  We assume that at $t=0$ the cavity has reached a steady state, at which time the drive is abruptly turned off.  Since the amplitude and phase at $t=0$ will depend on the detuning of the drive from the true resonant frequency while the cavity approached the steady state, for $t>0$ the average output voltage (on whichever port is to be measured) can be written
\begin{equation}
\langle V_{\mathrm{out}}(t>0) \rangle = V_{0}\left(\Delta_{0}\right) e^{-\kappa_{\mathrm{tot}}t/2} \cos\left[\omega_{0}(t)t + \phi(\Delta_{0})\right]
\end{equation}
where $\Delta_{0}$ is a function of $\delta\omega_{0}(t<0)$.  As we will see, the details of this dependence become irrelevant as long as care is taken in how the output signal is acquired and processed.  

Since microwave signals are generally too fast to be recorded directly, they are typically mixed down to an intermediate frequency before being recorded \cite{thesis_schuster}.  Here we consider a homodyne IQ detection scheme, where the output signal is mixed down to DC using a local oscillator at the drive frequency $\overline{\omega_{0}}$, after which both the in-phase (I) and quadrature (Q) components of the signal are recorded \cite{Silva2010, Eichler2012}.  In the presence of frequency fluctuations, these ostensibly DC signals now take the form
\begin{equation}
\begin{split}
I(t) &= V_{0}\left(\Delta_{0}\right) e^{-\kappa_{\mathrm{tot}}t/2} \cos\left[\delta\omega_{0}(t)t + \phi(\Delta_{0})\right] \\
Q(t) &= V_{0}\left(\Delta_{0}\right) e^{-\kappa_{\mathrm{tot}}t/2} \sin\left[\delta\omega_{0}(t)t + \phi(\Delta_{0})\right]
\end{split}
\end{equation}
which remain oscillatory due to $\delta\omega_{0}(t)$.  However, this oscillatory time dependence is not present in the amplitude of the signal
\begin{equation}
A(t) = \sqrt{I(t)^{2}+Q(t)^{2}} = V_{0}\left(\Delta_{0}\right) e^{-\kappa_{\mathrm{tot}}t/2}
\end{equation}
whose time dependence is a simple exponential decay from which the rate $\kappa_{\mathrm{tot}}/2$ is easily determined.  Even if the noise associated with the output signal is sufficiently large that an ensemble of measurements must be performed and averaged, only the initial value of the amplitude will be affected by this process such that $\kappa_{\mathrm{tot}}/2$ can still be determined from the rate of decay.  However, the ensemble-averaged amplitude will also acquire a DC offset due to the root mean square amplitude of the noise, which in practice could make it difficult to resolve the exponential decay.

To illustrate how the results of this measurement may be skewed by frequency fluctuations if the amplitude is not directly measured, we consider a single channel homodyne detection scheme in which only the in-phase part of the signal is recorded.  For simplicity, we assume that $V_{0}$ and $\phi$ are constants independent of $\Delta_{0}$, that $\phi=0$, and that the resonant frequency doesn't fluctuate appreciably over the course of a single measurement but does fluctuate over the course of repeated measurements.  In this case the in-phase part of the signal obtained from a single measurement takes the form
\begin{equation}
I(t) = V_{0} e^{-\kappa_{\mathrm{tot}}t/2} \cos(\delta\omega_{0}t).
\end{equation}
Assuming $\delta\omega_{0}$ is a Gaussian-distributed random variable with mean zero and variance $\sigma_{\omega_{0}}^{2}$, then the average time series from an ensemble of such measurements will take the form
\begin{equation}
\begin{split}
\langle I(t) \rangle &= V_{0} e^{-\kappa_{\mathrm{tot}}t/2} \int\limits_{-\infty}^{\infty}\cfrac{e^{-\Omega^{2}/2\sigma_{\omega_{0}}^{2}}}{\sqrt{2\pi\sigma_{\omega_{0}}^{2}}}\cos(\Omega t)d\Omega \\
&= V_{0} e^{-\kappa_{\mathrm{tot}}t/2} e^{-\sigma_{\omega_{0}}^{2}t^{2}/2}.
\end{split}
\end{equation}
Clearly, this signal is no longer ideal for extracting $\kappa_{\mathrm{tot}}$; if one tries to fit it to a simple exponential decay, then one will extract a total damping rate skewed by $\sigma_{\omega_{0}}$.  As before, this effect only becomes significant as $\sigma_{\omega_{0}}$ approaches $\kappa_{\mathrm{tot}}$.  

Thus, to accurately perform a ringdown measurement in the presence of frequency fluctuations, one should ideally detect the amplitude (or, equivalently, the power) of the signal directly.  As we've shown, this can be accomplished with an IQ measurement, but diode detection is another feasible approach.  How this is best accomplished, or if it can be accomplished at all, will depend on the specifics of the experimental device under consideration.  Regardless, if the signal to noise ratio of a single measurement is sufficiently small that an ensemble of measurements is required to resolve the signal, then it is absolutely necessary to compute the amplitude of the signal prior to averaging in order to avoid the results being skewed by frequency fluctuations.

\section{Time-domain Measurements of Frequency Fluctuations}

We now turn our attention to a common method for observing frequency fluctuations in the time domain \cite{Gao2007, Kumar2008, Barends2008, Barends2009}.  We again consider a continuous measurement of the scattering matrix elements, except instead of sweeping the detuning across the resonant frequency we keep it fixed at $\Delta =0$ and measure the quadratures of the scattered signal as a function of time.  For small fluctuations $\delta\omega_{0}(t)$ (relative to $\kappa_{\mathrm{tot}}$) about this detuning, we can approximate Eq. \eqref{eq:ideal_scattering_matrix_elements} as
\begin{equation}
S_{jk}(t) \approx \delta_{jk} - r_{jk}\left(1+e^{4i\delta\omega_{0}(t)/\kappa_{\mathrm{tot}}}\right) 
\end{equation}
such that the amplitude and phase of the scattered signal are given by
\begin{equation}
\begin{split}
|S_{jk}(t)| &\approx \sqrt{\delta_{jk} - 4\delta_{jk}r_{jk} + 4r_{jk}^{2}}\\
\mathrm{arg}\left(S_{jk}(t)\right) &\approx -i\log\left[\frac{\delta_{jk}-2r_{jk}}{|\delta_{jk}-2r_{jk}|}\right] -  \frac{4r_{jk}}{\delta_{jk}-2r_{jk}}\frac{\delta\omega_{0}(t)}{\kappa_{\mathrm{tot}}}
\end{split}
\end{equation}
to linear order in $\delta\omega_{0}(t)/\kappa_{\mathrm{tot}}$, valid for $\delta_{jk}\neq 2r_{jk}$.  We therefore see that frequency fluctuations are encoded as fluctuations in the scattered signal's phase to lowest order, while its amplitude remains constant.  Thus, if we can independently determine $\kappa_{\mathrm{tot}}$ and $r_{jk}$ then we can easily measure frequency fluctuations in the time domain by sampling the phase of the scattered signal as a function of time and inverting the above relationship.

For frequency fluctuations comparable to the cavity linewidth $\kappa_{\mathrm{tot}}$, however, the situation is more complicated for two key reasons.  First, we must use the techniques developed in the preceding sections to determine the parameters $\kappa_{\mathrm{tot}}$ and $r_{jk}$ using an appropriate model for $\overline{S_{jk}}$, otherwise our method for finding $\delta\omega_{0}(t)$ from $S_{jk}(t)$ will be inaccurate.  Second, fluctuations in the resonant frequency will now cause both the amplitude and phase of the scattered signal to vary, so we must measure both quantities as a function of time.  We can simplify the analysis of this time domain measurement of the scattered signal by translating our measurements $S_{jk}(t)$ along the real axis
\begin{equation}
\begin{split}
S_{jk}^{C}(t) &= S_{jk}(t) - \delta_{jk} + r_{jk} \\
&= -r_{jk}e^{-2i\arctan(-2\delta\omega_{0}(t)/\kappa_{\mathrm{tot}})}
\end{split}
\end{equation}
so that the center of the resonance circle lies at the origin of the complex plane.  Now, as before, fluctuations in the resonant frequency encode themselves as fluctuations in the phase of the scattered signal according to
\begin{equation}
\mathrm{arg}\left(S_{jk}^{C}(t)\right) = -2\arctan(-2\delta\omega_{0}(t)/\kappa_{\mathrm{tot}})
\end{equation}
which can once again be inverted easily to find $\delta\omega_{0}(t)$.  We note that this measurement can be used to corroborate the random-variable model for $\delta\omega_{0}$ by creating a histogram from the time series of $\delta\omega_{0}(t)$ and comparing to the expected probability distribution $P(\Omega)$.  

The primary strength of this time domain measurement is that it allows one to measure the power spectral density of frequency fluctuations, rather than simply treating them as random variables with an effective probability distribution.  However, it also has a major downside: it is only sensitive to those fluctuations occurring on time-scales slower than the time it takes to sample the quadratures of the scattered signal, which is the exact opposite sensitivity range as that of the frequency-domain method we have presented.  Since this sampling rate is usually set by the bandwidth of the measurement, faster sampling means more measurement noise.  This must be reduced far below the level of the phase fluctuations of interest in order to resolve the effect of frequency fluctuations, which in practice means operating at either high powers or long time scales.  Often it is either desirable \cite{Burnett2018} or necessary \cite{Krantz2013} to perform this characterization at the single-photon level, which means this measurement will only be sensitive to very slow noise processes.  

This may suffice in many cases, most notably when the frequency fluctuations have a $1/f$ spectrum.  This has been shown to be the dominant spectral component of the intrinsic frequency fluctuations present in microwave cavities, which arise due to fluctuating two-level systems in the dielectric substrates on which planar cavities are fabricated \cite{Gao2007, Barends2008, Barends2009, Burnett2014, Burnett2018}.  It is also likely to be dominant for cavities tuned by the flux threading SQUID loops \cite{Palacios-Laloy2008, Sandberg2008, Krantz2013, Lin2014, Krantz2016, Svensson2017, Bengtsson2018, Svensson2018} due to the $1/f$ flux noise ubiquitous in SQUIDs \cite{Wellstood1987, Sendelbach2008, Kumar2016}, which is believed to arise from unpaired surface spins.  We do not expect to resolve the frequency fluctuations due to the Kerr nonlinearity using this method, however, since these fluctuations are likely to occur far faster than the measurement times required at the single photon level.  Further complicating this measurement when fast fluctuations are present is the fact that each sample of the scattering matrix element $S_{jk}$ obtained will correspond to an average taken over all the faster fluctuations.  So, rather than sampling from $S_{jk}$ one will actually be sampling from $\overline{S_{jk}}$.  

Lastly, it is worth noting that some of these difficulties may be alleviated by using a Pound-locking loop \cite{Lindstroem2011, Burnett2013, Burnett2014, Graaf2014, Burnett2018}, but this technique has thus far only been applied in the case of small frequency fluctuations.  As such, it may have to be extended in order to study frequency fluctuations in tunable and nonlinear microwave cavities.  Furthermore, this technique requires specialized equipment, whereas one can measure the scattering matrix elements using standard equipment (typically a vector network analyzer) present in every lab studying microwave-frequency devices.

\section{Discussion}

In conclusion, we have shown how frequency fluctuations comparable to the cavity linewidth may arise in tunable and nonlinear microwave cavities, and have presented a model for how these fluctuations deform the trajectories traced out by the scattering matrix elements in the complex plane.  For tunable cavities we have shown how fluctuations in the tuning parameter induce fluctuations in the resonant frequency, and that failing to account for these fluctuations may lead one to extract damping rates that appear to vary with the tuning parameter, which is a common observation in these systems.  For Kerr cavities we have shown that quantum fluctuations in the cavity quadratures induce frequency fluctuations, which can appreciably affect the apparent damping rates of the cavity as the strength of the nonlinearity approaches the cavity linewidth.  However, by using the model we have presented for the average scattering matrix elements as a fitting function for experimental data, one can extract the true damping rates even in the presence of these frequency fluctuations, allowing these cavities to be characterized more accurately moving forward.  Further, we have discussed how ringdown measurements can be performed in the presence of frequency fluctuations so as to accurately extract the total damping rate $\kappa_{\mathrm{tot}}$, which can serve to corroborate the true damping rates of cavities.  Finally, we have extended a standard method for measuring frequency fluctuations in the time domain to the case of fluctuations comparable to the cavity linewidth, and compared these two measurement paradigms.  

It is straightforward to extend our model to many other systems that we have not considered explicitly, but for which the analysis will be similar to that presented here.  First and foremost, the convolution of Eq. \eqref{eq:average_scattering_matrix_element} can be used for any scattering matrix elements $S_{jk}(\Delta)$ and any probability distribution $P(\Omega)$.  Since most $S_{jk}$'s are similar in form to those of the two-sided cavity, we expect similar closed-form expressions exist for this convolution in the case of the probability distributions we've considered.  For other probability distributions, this convolution can be implemented numerically.  Our treatment can also be extended easily to the case of impedance mismatches, the effect of which is generally a rotation of the resonance circle around the off-resonant point ($S_{jk}=\delta_{jk}$ for a two-sided cavity) and an expansion of its radius \cite{Khalil2012,Megrant2012}, transformations which can be induced on $S_{jk}$ by the addition and multiplication of complex numbers.  Since these operations commute with the convolution, they can be applied to the average scattering matrix elements $\overline{S_{jk}}$ in exactly the same way.

For the case of tunable cavities, our treatment is easily extended to the case of multiple tuning parameters.  If $\omega_{0} = \omega_{0}(x_{1},...,x_{N})$ and fluctuations in each of the $x_{i}$ are independent, then Eq. \eqref{eq:tunable_variance} becomes
\begin{equation}
\sigma_{\omega_{0}}^{2} = \sum\limits_{i=1}^{N}\left|\frac{\partial\omega_{0}(\overline{x_{1}},...,\overline{x_{N}})}{\partial x_{i}}\right|^{2}\sigma_{x_{i}}^{2}
\end{equation}
to linear order in the fluctuations $\delta x_{i}$, where $\sigma_{x_{i}}^{2}$ is the variance of fluctuations in $\delta x_{i}$.  We may also apply the same techniques we've developed to any dispersive coupling to the cavity, not just the classical tuning parameter considered here.  Common examples are the dispersive Jaynes-Cummings interaction between the cavity and a qubit, the optomechanical interaction between the cavity and a mechanical resonator, and the cross-Kerr interaction between the cavity and either another cavity or another mode of the same cavity.  In these cases, fluctuations in the state of the coupled system will cause the resonant frequency of the cavity to fluctuate in the same manner discussed here, leading to analogous effects.

For the case of the Kerr cavity, it may be possible to extend our treatment into the regime of single-photon strength for which $|K|\gtrsim\kappa_{\mathrm{tot}}$.  To do so, one would have to treat the steady-state dynamics beyond the semi-classical approximation, using either numeric \cite{Casteels2016, Braasch2019} or analytic techniques \cite{Drummond1980, Bartolo2016}.  However, such systems may simply require the use of different characterization techniques than those considered here, rendering this extension moot.  A more straightforward extension is to consider a finite temperature, in which case the steady state will be a displaced thermal state.  Fluctuations in the quadratures will once again be Gaussian but they will now have a variance $1+2n_{\mathrm{th}}$, where
\begin{equation}
n_{\mathrm{th}} = \cfrac{1}{e^{\hbar\omega_{0}/kT}-1}
\end{equation}
is the average thermal occupation of the cavity.  All the results obtained in Sections \ref{sec:kerr_cavity} and \ref{sec:tunable_kerr_cavity} can be extended to this case by scaling the Kerr nonlinearity by this new variance according to $K\rightarrow K(1+2n_{\mathrm{th}})$.

Our model can also be applied to the intrinsic frequency fluctuations present in microwave cavities due to the fluctuating two-level systems in dielectric substrates \cite{Gao2007, Kumar2008, Barends2008, Barends2009, Lindstroem2011, Burnett2013, Burnett2014, Graaf2014, Burnett2018}.  As improved fabrication techniques are developed to reduce damping rates in these cavities, the scale of intrinsic frequency fluctuations may become comparable to the cavity linewidths, yielding appreciable deformation of the scattering matrix elements.  However, the damping rates in these cavities are usually limited by the presence of the same two-level systems responsible for intrinsic frequency fluctuations \cite{Wang2009, Barends2010}.  There is considerable experimental effort to understand, characterize, and reduce the effects of two-level systems in microwave cavities \cite{Megrant2012, Geerlings2012, Bruno2015, Woods2019}, due in large part to the detrimental effect they have on the coherence times of superconducting qubits \cite{Martinis2005, Gunnarsson2013}.  It remains to be seen whether these efforts will have a significant effect on the ratio $\sigma_{\omega_{0}}/\kappa_{\mathrm{tot}}$ for intrinsic frequency fluctuations.  Finally, our model can be extended to account for intrinsic fluctuations of the cavity damping rates \cite{Graaf2014, Earnest2018} by treating these parameters as random variables and performing a second convolution with respect to them.  As in all previous cases, the scale of these fluctuations would have to be comparable to the total damping rate to yield an appreciable effect.

\begin{acknowledgments}
We would like to thank B. Thyagarajan, S. Kanhirathingal, W. Braasch, and C. Ramanathan for helpful comments and discussions.  B.~L.~B. and A.~J.~R. were supported by the National Science Foundation under Grant No. DMR-1807785.  M.~P.~B. was supported by the National Science Foundation under Grant No. DMR-1507383.
\end{acknowledgments}


%

\end{document}